\documentclass[11pt]{article}             
\usepackage{osid} %
\usepackage{geometry}
\usepackage{graphicx}
\usepackage{amssymb}
\usepackage{amsmath}
\usepackage{epstopdf}
\usepackage{cite}
\usepackage[sans]{dsfont}
\usepackage[mathscr]{euscript}
\newcommand{\rd}{\mathrm{d}}
\newcommand{\ri}{\mathrm{i}}
\newcommand{\re}{\mathrm{e}}
\newcommand{\openone}{\mathds1}
\newcommand{\norm}[1]{\left\Vert#1\right\Vert}
\newcommand{\abs}[1]{\left\vert#1\right\vert}

\newcommand{\Pbb}{\mathbb{P}}
\newcommand{\Qbb}{\mathbb{Q}}
\newcommand{\Ebb}{\operatorname{\mathbb{E}}}
\newcommand{\Cov}{\operatorname{Cov}}
\newcommand{\Var}{\operatorname{Var}}
\newcommand{\RE}{\operatorname{Re}}

\newcommand{\Tr}{\operatorname{Tr}}

\newcommand{\Fscr}{\mathscr{F}}
\newcommand{\Hscr}{\mathscr{H}}
\newcommand{\Lcal}{\mathcal{L}}
\newcommand{\Rcal}{\mathcal{R}}

\newcommand{\Xcal}{\mathcal{X}}
\newcommand{\Vcal}{\mathcal{V}}

\renewcommand{\thesection}{\arabic{section}}
\renewcommand{\thesubsection}{\thesection.\arabic{subsection}}

\renewcommand{\RHD}{I. Semina, V. Semin, F. Petruccione, A. Barchielli \qquad \qquad Markovian and non-Markovian SSEs}

\title{Stochastic Schr\"{o}dinger Equations for Markovian and non-Markovian cases }

\author{I.~Semina\footnote{\ \ iuliia.semina@mail.ru}, V. Semin\footnote{\ \ semin@ukzn.ac.za}, F. Petruccione\footnote{\ \ also  National Institute for Theoretical Physics (NITheP) -- petruccione@ukzn.ac.za}, \\{\footnotesize\ Quantum Research Group, School of Chemistry and Physics,}\\ {\footnotesize\
        University of KwaZulu-Natal, 4001, South Africa,} \\
and A.~Barchielli\footnote{\ \ also Istituto Nazionale di Fisica Nucleare (INFN), Sezione di Milano, \\ \indent and Istituto Nazionale di Alta Matematica (INDAM-GNAMPA) -- alberto.barchielli@polimi.it}\\ {\footnotesize\ Politecnico di Milano, Dipartimento di Matematica,} \\
{\footnotesize\ piazza Leonardo da Vinci 32, 20133 Milano, Italy.}}

\begin{document}

\maketitle
\begin{abstract}
Firstly, the Markovian stochastic Schr\"{o}dinger equations are presented, together with their connections with the theory of measurements in continuous time. Moreover, the stochastic evolution equations are translated into a simulation algorithm, which is illustrated by two concrete examples --- the damped harmonic oscillator and a two-level atom with homodyne photodetection. Then, we consider how to introduce memory effects in the stochastic Schr\"{o}dinger equation via coloured noise. Specifically, the approach by using the Ornstein-Uhlenbeck process is illustrated and a simulation for the non-Markovian process proposed. Finally, an analytical approximation technique  is tested with the help of the stochastic simulation in a model of a dissipative qubit.
\end{abstract}

\section{Introduction}
Typically, an open quantum system is a system interacting with an external environment which experimentalists cannot control \cite{Dav76,4,28}. It is well known that the dynamics of an open quantum system can be described in one of the following ways: local and non-local master equations for the density matrix \cite{37,38,39}, Feynman's path integrals \cite{40}, stochastic Schr\"{o}dinger equations (SSE) \cite{7,28,27,Bel89,last} and quantum trajectories \cite{7,28}. In this review we give a description of the technique based on the SSE, which can be useful for the description of Markovian and non-Markovian dynamics of open quantum systems. Moreover, we shall illustrate the Markovian and non-Markovian theory by giving some simulations.
In the non-Markovian case we also use the stochastic simulations to check the validity of an analytic approximation for the mean state.

The stochastic representation of quantum Markovian processes already appeared in the fundamental work by Davies \cite{Dav76,19} and it was applied to the derivation of a photocounting formula. While the theory was originally formulated in terms of a stochastic process for the reduced density matrix, it was recognized by Barchielli and Belavkin \cite{7}, Dalibard, Castin and M\o{}lmer \cite{18} and by Dum, Zoller and Ritsch \cite{17} that it can also be formulated as a stochastic process for the state vector in the reduced system Hilbert space and that it leads to efficient numerical simulation algorithms. At the same time, there has been considerable interest in the unravelling of master equations for density operators into quantum trajectories which are the realizations of the underlying stochastic process \cite{28}. Just as different ensembles of state vectors may be represented by one density operator, one master equation may be decomposed in many different ways into SSEs.

The SSE is a differential equation for a wave-function process $\psi(t)$ which contains a stochastic term to describe the relaxation dynamics of an open quantum system. The link with the traditional master equation is given by the average property $\Ebb[|\psi(t)\rangle\langle\psi(t)|] = \eta(t)$, where $\Ebb$ denotes the ensemble average over the realizations of process $\psi(t)$ and $\eta(t)$ is the statistical operator satisfying the master equation. To find the SSE providing a given master equation by averaging is called \textit{unravelling} \cite{1}.

Also, in special situations, the SSE can be interpreted in terms of quantum measurements. In these cases, the solution $\psi(t)$ is called a \textit{quantum trajectory} \cite{11} and describes the evolution of an open system undergoing indirect continuous measurement. This interpretation is important for understanding quantum optics experiments such as direct photo-detection, spectral photo-detection, homodyning and heterodyning \cite{12,13,14,3}.

In the regime of the validity of the Markov approximation (no memory effects) \cite{26} it is known how to construct an appropriate unravelling in terms of a SSE. It is always possible to derive a linear SSE for a non-normalized vector $\phi(t)$, such that $\psi(t)=\norm{\phi(t)}^{-1}\phi(t)$. Moreover, the linear and nonlinear versions of the SSE are related by a change of probability measure, and it is this link that allows for a measurement interpretation \cite{3}. Also, these stochastic differential equations can be deduced from purely quantum evolution equations for the measured system coupled with a quantum environment, combined with a continuous monitoring of the environment itself \cite{29,Bar06,BarG13}.

In the non-Markovian case \cite{23,21,22}, to find relevant SSEs describing both non-Markovian quantum evolutions and continuous monitoring is a complex task. Other than in the Markovian case, no general theory has been developed. Nevertheless, it is possible to follow a general strategy. This strategy is first to generalize directly the Markovian SSE, second to show if it provides an unravelling of a corresponding master equation, and third to check if it has a measurement interpretation \cite{1,BarH95,2,apr,BarG12}. To work at the Hilbert space level guarantees automatically the complete positivity of the evolution of the statistical operator.  It seems possible to adapt the Markovian approach by replacing white noises with more general noises and by allowing for random coefficients in the equation. We will show how to introduce memory effects in the SSE with the help of coloured noise. Specifically, we will illustrate the approach by replacing the Wiener process with the  Ornstein-Uhlenbeck process. Such approaches are efficient for simulating corresponding non-Markovian evolutions. Also, the non-Markovian SSE is formulated in a way that allows for an interpretation in terms of measurements in continuous time.

The paper introduces the general theory of the SSE as well as the corresponding simulation techniques and is structured as follows. Section \ref{sec:SSE} describes the general theory of the SSEs in the Markovian case. It presents the general mathematical framework of the linear and nonlinear SSE. We consider a linear stochastic equation with ``multiplicative noise'' for the wave function $\phi(t)$ in the purely diffusive case. Then, we discuss how to get the physical probabilities and we derive the nonlinear SSE for the conditional states $\psi(t)$. In Section \ref{sec:simM} we describe the simulation techniques for SSEs and we show the simulations for two Markovian processes, the damped harmonic oscillator and a two-level atom with homodyne photodetection.  Section \ref{sec:memory} is devoted to the introduction of coloured noise in the SSEs; we limit the presentation of this part of the theory to a restricted, but significant, class of SSEs with memory. The simulation of such non-Markovian processes is also proposed and applied as a test of other approximation techniques. In Section \ref{sec:concl} we briefly summarize the main results and indicate some directions of future work. Basic concepts from the theory of stochastic processes are summarized in the Appendix \ref{appen}.

\section{Stochastic Schr\"{o}dinger equations}\label{sec:SSE}
In this paper we will show the approach to the theory of open quantum systems based on stochastic differential equations (SDEs), with particular emphases on continuous measurements. In this theory there are four kinds of SDEs: the linear stochastic Schr\"{o}dinger equation (lSSE), a linear SDE for non-normalized vectors in the Hilbert space of the system \eqref{2.1.3}, the SSE, a nonlinear SDE for normalized vectors in the Hilbert space \eqref{SSE1}, the linear stochastic master equation \cite[Sections 3.1.2, 3.4.1]{3}, a linear SDE for positive trace-class operators, and the stochastic master equation \cite[Sections 3.5, 5.1]{3}, a nonlinear SDE for density matrices. Two kinds of noises may appear in the SSEs and characterize the jump and the diffusive cases. Here we will focus on the diffusive case. For SSEs and SMEs of the diffusive type a Wiener process $B$ appears in the linear equations and a Wiener process $W$ in the nonlinear equations; $B$ and $W$ are connected by the Girsanov transformation \eqref{newW}.

To have some hints on what we will construct, let us consider an instantaneous and pure state preserving measurement of some quantity $X$ with discrete values $\{x_k\}$. In the Hilbert space formulation of quantum mechanics, such an observation is represented by a collection of operators $\{E_k\}$ such that $\sum_kE_k^{\;\dag}E_k=\openone$; these operators acts on $\Hscr$, the Hilbert space of the system. The map $x_k \mapsto E_k^{\;\dag}E_k$ is a (discrete) positive-operator valued measure, the modern generalization of quantum observable. Let $\varphi\in \Hscr$, $\norm{\varphi}=1$, be the pre-measurement state and set $\phi_k=E_k\varphi$. Then, $\norm{\phi_k}^2\equiv \langle \varphi|  E_k^{\;\dag}E_k\varphi\rangle$ is interpreted as the probability of the result $\{X=x_k\}$ in the measurement and $\psi_k=\phi_k/\norm{\phi_k}$ as the state of the system after the measurement given the result $\{X=x_k\}$. The conditional state $\psi_k$ is often called the \textit{a posteriori state} \cite{3}. For the case of measurement in continuous time the output is not discrete, but it is a whole trajectory of some observed quantity; this brings into play the stochastic processes. Apart from this complication, the lSSE is an evolution equation for the analog of the non-normalized vectors $\phi_k$, while the SSE is the evolution equation for the analog of the post-measurement states $\psi_k$. Note that the map $\varphi\mapsto \phi_k=E_k\varphi$ is linear, while the map $\phi\mapsto \psi_k$ is non-linear due to the normalization; the same difference will characterize the passage from the lSSE to the SSE.

\subsection{The linear stochastic Schr\"odinger equation}\label{sec:lSSE}

The SDEs we consider are driven by white noise. Some notions on Wiener process and stochastic calculus are given in Appendix \ref{appen}, but for a full presentation see \cite{KarS91, Mao97} and for a summary see \cite{3}.

First of all we work in a reference probability space 
($\Omega, \Fscr, \Qbb$), where $\Omega$ is the sample  space, $\Fscr$ the $\sigma$-algebra of events, and $\Qbb$ a reference probability. A filtration is a family $(\Fscr_t)_{t\geqslant  0}$ of increasing sub-$\sigma$-algebras of $\Fscr$, i.e. $\Fscr_s\subset \Fscr_t\subset \Fscr$ for $0\leq  s< t< +\infty$. Sometimes, $(\Omega, \Fscr, (\Fscr_t), \Qbb)$ is said to be a stochastic basis. Typically, a filtration describes the accumulation of information during time: each $\Fscr_t$ is the collection of all the events which we can decide whether they have been verified or not by observations up to time $t$. In the basis $(\Omega, \Fscr, (\Fscr_t), \Qbb)$ a continuous, adapted $d$-dimensional Wiener process  $B=\{B_j(t),\; t\geq  0,\; j=1,\ldots,d\}$ is defined (see Appendix \ref{Wiener}).

Let us start from a generic homogeneous linear SDE with ``multiplicative'' noise for the process $\phi(t)$ \cite{3}:
\begin{equation}\label{2.1.1}
\rd\phi(t) = K(t)\phi(t)\,\rd t +\sum_{j=1}^{d} R_j(t)\phi(t)\,\rd B_j(t),
\end{equation}
where $\phi(0)=\psi_0$, $\psi_0\in \Hscr$, the coefficients $R_j(t), K(t)$ are (non-random) linear operators on $\Hscr$. The SDE \eqref{2.1.1} is to be intended in integral sense and the solution  $\phi$ is the continuous, adapted It\^{o} process satisfying
\[
\phi(t)=\psi_0+\int_{0}^{t}K(s)\phi(s)\,\rd s+\sum_{j=1}^{d}\int_{0}^{t}R_j(s)\phi(s)\,\rd B_j(s).
\] 
The last term in the above equation is a stochastic It\^{o} integral (see Appendix \ref{Int}).

\paragraph{The physical probability.}
To develop the theory, we need  $\norm{\phi(t)}^2$ to be a probability density, cfr.\ the hints at the beginning of Section \ref{sec:SSE}. Precisely, let us define
\begin{equation}\label{newprob}
\Pbb^t_{\psi_0}(F):=\int_F \norm{\phi(t,\omega)}^2\,\Qbb(\rd\omega)=\Ebb_\Qbb[\norm{\phi(t)}^21_F], \quad \forall F\in \Fscr_t,
\end{equation}
where $1_F$ is the indicator function of the set $F$. To guarantee that \eqref{newprob} defines a probability measure, we have to ask only the normalization: \begin{equation}\Ebb_\Qbb[\norm{\phi(t)}^2]=1, \qquad \forall t\geq 0.\end{equation}
Obviously observations in the future cannot change the probabilities on past events and to get this we need a consistency property: \begin{equation}\label{consistency}
\Pbb^t_{\psi_0}(F)=\Pbb^s_{\psi_0}(F), \qquad \forall F\in \Fscr_s, \qquad \forall t,s:t\geq s \geq 0.
\end{equation}
This is equivalent to asking $\norm{\phi(t)}^2$ to be a $\Qbb$-\emph{martingale}  (Appendix \ref{Martingale}). Then, its mean is a constant and the normalization for every time reduces to the normalization of the initial state $\psi_0$.

Using It\^{o}'s lemma (Appendix \ref{Calc}) for $\rd\Arrowvert\phi(t)\Arrowvert^2$  we can derive as in \cite[Section 2.2.3]{3}:
\begin{multline}\label{2.1.2}
\Arrowvert\phi(t)\Arrowvert^2=\Arrowvert\psi_0\Arrowvert^2+\int_{0}^{t}\big\langle\phi(s)\big|\biggl( K(s)+K(s)^{\dag}+\sum_{j}R_j(s)^{\dag}R_j(s)\biggr)\phi(s)\big\rangle\,\rd s  \\
{}+\sum_{j=1}^{d}\int_{0}^{t}\big\langle\phi(s)\big|\left(R_j(s)+R_j(s)^{\dag}\right)\phi(s)\big\rangle\,\rd B_j(s).
\end{multline}
In order to reduce $\Arrowvert\phi(t)\Arrowvert^2$ to a martingale, we need the integrand in the time integral in Eq.\  (\ref{2.1.2}) to vanish for every initial condition, i.e.
\begin{equation}
K(t)+K(t)^{\dag}+\sum_jR_j(t)^{\dag}R_j(t)=0.\nonumber
\end{equation}
Then, the operator $K(t)$ has the structure
\begin{equation}\label{K(t)}
K(t)=-\ri H(t)-\frac 1 2 \sum_{j=1}^{d}R_j(t)^{\dag}R_j(t),
\end{equation}
where $H(t)$ is a self-adjoint operator on $\Hscr$, called the \textit{effective Hamiltonian} of the system.

\paragraph{The lSSE.} Finally, the \textit{linear stochastic Schr\"{o}dinger equation} (diffusive type) is given by
\begin{gather}\label{2.1.3}
\rd \phi(t)=\biggl(-\ri H(t)-\frac{1}{2}\sum_{j=1}^{d}R_j(t)^{\dag}R_j(t)\biggr)\phi(t)\,\rd t+\sum_{j=1}^{d}R_j(t)\phi(t)\,\rd B_j(t),
\\
\phi(0)=\psi_0, \quad \psi_0\in\Hscr, \quad \norm{\psi_0}=1, \qquad H(t)=H(t)^{\dag}.
\end{gather}

The linear stochastic Schr\"{o}dinger equation \eqref{2.1.3} reduces to an ordinary Schr\"{o}dinger equation $\rd\phi(t)/\rd t =-\ri H(t)\phi(t)$ when we switch off the measurement and the interactions with the environment $(R_j(t) \equiv 0)$.

\subsection{The a posteriori states, the output and the master equation}

Let us consider now a finite time interval $[0,T]$; the current time $t$ will be always inside this interval. We also introduce the normalized version $\psi(t)$ of the vector $\phi(t)$:
\begin{equation}\label{apos}
\psi(t)= \frac{\phi(t)}{\norm{\phi(t)}}.
\end{equation}

Then, the interpretation of the theory is similar to the hints given at the beginning of Section \ref{sec:SSE} and it is given here below.
\begin{enumerate}
\item The physical probability of the events occurring up to time $T$ is $\Pbb^T_{\psi_0}$. By the consistency property \eqref{consistency} the choice of $T$ is immaterial.
\item The cumulated output of the continuous measurement is the $d$-dimensional process $B$ and its distribution is given by the physical probability, so that it is no more a Wiener process. More precisely the output in any time interval $[s,t]$ is $B(t)-B(s)$, so that the instantaneous output is the formal time derivative $\dot B(t)$. The structure of the output under the physical probability is given in Eq.\ \eqref{Gir}.
\item The normalized vector $\psi(t)$ \eqref{apos} is the a posteriori state, i.e.\ the conditional state of the system at time $t$ given the observed output up to time $t$. The evolution of $\psi(t)$ is given by the SSE \eqref{SSE1}.
\end{enumerate}

Let us introduce now the average state
\begin{equation}\label{eta(t)1}
\eta(t)=\Ebb_{\Pbb^T_{\psi_0}} [ |\psi(t)\rangle\langle \psi(t)| ] \equiv \int_\Omega |\psi(t,\omega)\rangle\langle \psi(t,\omega)|  \Pbb^T_{\psi_0}(\rd \omega), \qquad T\geq t\geq 0.
\end{equation}
Note that, by construction, $\eta(t)$ is a positive operator and that, by the normalization of $\psi(t)$, one has $\Tr\{\eta(t)\}=1$, so that $\eta(t)$ is a statistical operator.
\begin{enumerate}\setcounter{enumi}{3}
\item The statistical operator $\eta(t)$ is the state we attribute to the system at time $t$, when the output is not known;  it is called the \emph{a priori state} and satisfies the master equation \eqref{master}.
\end{enumerate}

By the consistency property \eqref{consistency}  we can take $T=t$. Then, by the fact that $\Pbb^T_{\psi_0}(\rd \omega)=\norm{\phi(t,\omega)}^2 \Qbb(\rd \omega)$ and $\norm{\phi(t,\omega)}^2 |\psi(t,\omega)\rangle\langle \psi(t,\omega)| = |\phi(t,\omega)\rangle\langle \phi(t,\omega)| $, we get the equivalent expression
\begin{equation}\label{eta(t)2}
\eta(t)=\Ebb_\Qbb [ |\phi(t)\rangle\langle \phi(t)| ]\equiv  \int_\Omega |\phi(t,\omega)\rangle\langle \phi(t,\omega)|  \Qbb(\rd \omega) .
\end{equation}

By computing the stochastic differential of $|\phi(t)\rangle\langle \phi(t)|$ and by taking the $\Qbb$-mean of the resulting equation one gets the \emph{master equation}
\begin{subequations}\label{master}
\begin{equation}
\dot \eta(t) = \Lcal(t)[\eta(t)],
\end{equation}
\begin{equation}\label{L(t)}
\Lcal(t)[\varrho]= -\ri [H(t),\varrho] + \sum_{j=1}^d \biggl(R_j(t)\varrho R_j(t)^\dag - \frac 1 2 \left\{R_j(t)^{\dag}R_j(t),\varrho\right\}\biggr).
\end{equation}
\end{subequations}
Note that the Liouville operator $\Lcal(t)$ turns out to be in the usual Lindblad form.

From Eq.\ \eqref{2.1.2} with condition \eqref{K(t)} and the normalization of $\psi_0$, we get \cite[Section 2.3.1]{3}, by the rules of stochastic calculus,
\begin{equation}\label{norm2phi}
\norm{\phi(t)}^2= \exp\biggl\{\sum_j\int_0^t m_j(s)\,\rd B_j(s) -\frac 1 2 \int_0^t m_j(s)^2\, \rd s\biggr\},
\end{equation}
\begin{equation}\label{mj}
m_j(t)=2 \RE  \langle \psi(t)|R_j(t) \psi(t) \rangle.
\end{equation}
Then, Girsanov theorem gives that under the probability $\Pbb^T_{\psi_0}$ the process
\begin{equation}\label{newW}
W_j(t)=B_j(t) - \int_0^t m_j(s)\,\rd s, \qquad j=1,\ldots, d, \quad t\in[0,T],
\end{equation}
is a $d$-dimensional Wiener process \cite[Sections 2.3.2 and A.5.4]{3}.

Obviously we can write
\begin{equation}\label{Gir}
B_j(t)=W_j(t)+\int_{0}^{t}m_j(s) \rd s.
\end{equation}
Then, we can say that the instantaneous output $\dot B_j(t)$ is the sum of the white noise $\dot W_j(t)$ and the regular process $m_j(t)$ (the signal). However, let us stress that white noise and signal are not in general independent under the physical probability.

The theory of continuous measurements gives also all the correlations of the output process \cite[Section 4.3]{3}. In particular, by taking the mean of both sides in Eq.\ \eqref{Gir} and by taking into account Eqs.\  \eqref{eta(t)1} and \eqref{mj}, we get immediately the mean of the output
\begin{equation}\label{Eout}
\Ebb_{\Pbb^T_{\psi_0}}[B_j(t)]=\Tr\left\{\left(R_j(t)+R_j(t)^\dagger\right)\eta(t)\right\}.
\end{equation}
This equation suggests to interpret the $j$-th output as a continuous indirect monitoring of the system quantum observable $R_j(t)+R_j(t)^\dagger$. However, the final interpretation depends on the specific model. The output $B_j$ could also represent the photocurrent in homodyne or heterodyne detection; in this case  the system operator $R_j(t)$ depends on the interaction with the electromagnetic field and on the local oscillator wave. The channel $j$ could also represent a pure dissipative effect due to the environment; in this case $B_j(t)$ is not observed and the role of this channel is only for introducing a dissipative contribution into the Liouville operator \eqref{L(t)}.

\subsection{The nonlinear stochastic Schr\"odinger equation}\label{sec:nlSSE}
By using the rules of It\^o calculus and the lSSE, it is possible to compute the stochastic differential of the a posteriori state  $\psi(t)=\norm{\phi(t)}^{-1}\phi(t)$. By expressing the result in terms of the new Wiener process \eqref{newW}, the final result is the SSE
\begin{multline}\label{SSE1}
\rd\psi(t)=\sum_j\biggl[R_j(t)- \frac12\,  m_j(t)\biggr]\psi(t)\,\rd W_j(t)
\\ {}+\biggl[-\ri H(t) - \frac 12 \sum_j R_j(t)^\dag R_j(t)+\frac 1 2 \sum_jm_j(t)R_j(t)-\frac{1}{8}\sum_j m_j(t)^2\biggr]\psi(t)\,\rd t.
\end{multline}
As $m_j(t)$ \eqref{mj} is a bilinear function of $\psi(t)$, the SSE \eqref{SSE1} turns out to be a closed SDE for the process $\psi(t)$ under the probability $\Pbb^T_{\psi_0}$ \cite[Section 2.5.1]{3}.

Let us note that the master equation \eqref{master} is invariant under the transformation $R_j(t) \to \re^{\ri \theta_j}R_j(t)$. However, this is not true for the lSSE \eqref{2.1.3}, the SSE \eqref{SSE1} and the output \eqref{Gir}; indeed, $m_j(t)$ \eqref{mj} and its mean \eqref{Eout} are sensible to the phase of $R_j(t)$. So, the a posteriori states and the output depend on a phase shift in the operators of the dissipative part, while the mean dynamics is independent from such phases.


It is possible also to start from the SSE \eqref{SSE1}. In this case $W$ is a Wiener process under a probability $\Pbb$, which is directly the physical probability. Then, the output is defined by Eqs.\ \eqref{Gir} and \eqref{mj} and a lSSE can be introduced by a change of normalization and of probability \cite[Section 2.5.4]{3}.
A characteristic feature of the non-linear SSEs is to preserve the normalization of the state $\psi(t)$.

\subsection{The case of a random unitary evolution}\label{sec:randomunitary}
A very particular case is when all the operators $R_j(t)$ are anti-selfadjoint:
\begin{equation}
R_j(t) = -\ri V_j(t),\qquad V_j(t)^{\dag} = V_j(t).
\end{equation}
Then, Eqs.\ \eqref{newprob}, \eqref{apos}, \eqref{norm2phi}, \eqref{mj}, \eqref{Gir} give $m_j(t)=0$, $\norm{\phi(t)}^2=1$, $\psi(t)=\phi(t)$, $\Pbb^T_{\psi_0}=\Qbb$, $W_j(t)=B_j(t)$. This means that the $B_j$ are pure noises and there is no true measurement on the system. Moreover, the lSSE and the nonlinear one coincide and give a random unitary evolution:
\begin{equation}\label{Uevol}
\rd\psi(t)=-\ri\biggl[H(t)\,\rd t+\sum_{j}V_j(t)\,\rd W_j(t)\biggr]\psi(t)-\frac{1}{2}\sum_{j}V_j(t)^2\psi(t)\,\rd t.
\end{equation}
Formally, $H(t)+\sum_{j}V_j(t)\,\dot W_j(t)$ is the random Hamiltonian which generate the the unitary evolution. The last term is the It\^o correction due to the presence of the white noise $\dot W_j(t)$ in the formal Hamiltonian.
This class of SSEs was introduced as a model of dissipative evolution, without observation. In this case, all the physical quantities are obtained as a mean with respect to $W$ \cite{35}.

\section{Simulating SSEs for the Markovian case}\label{sec:simM}
The idea of unravelling has been a real breakthrough for simulating master equations; it is at the root of the Monte-Carlo wave function method \cite{4,23,30,31}. The basic idea of these methods is to generate independent realizations of the underlying stochastic process by a numerical algorithm and to estimate with the help of statistical means all desired expectation values from a sample of such realizations. A stochastic simulation thus amounts to perform an experiment on a computer. It yields the outcomes of single runs with their correct probabilities and provides, in addition to the mean values, estimates for the statistical errors of the quantities of interest.

Let us consider the SSE \eqref{SSE1} for the a posteriori states $\psi(t)$, with a standard Wiener process $W$ in a stochastic basis $\bigl(\Omega, \Fscr, (\Fscr_t), \Pbb\bigr)$.

A stochastic simulation algorithm serves to generate a sample of independent realizations of the stochastic process $\psi(t)$ for the conditional wave function. Let us denote these realizations by $\psi^r(t)$, $r=1,2,...,R$, where $R$ is the number of realizations in the sample.
A quantity of interest can be thought as a real functional $F[\psi,t]$ of the a posteriori states $\psi(s)$, $s\in[0,t]$; then, let
\begin{equation}\label{Mt}
M_t=\Ebb_\Pbb\big[F[\psi,t]\big]
\end{equation}
be its mean value.
An \emph{unbiased and consistent estimator} for the expectation value $M_t$ is provided by the \emph{sample mean}
\begin{equation}
\widehat{M}_t=\frac{1}{R}\sum_{r=1}^{R}F[\psi^r,t],
\end{equation}
where a hat is used to indicate an estimator. It is clear that the estimate is subjected to statistical  errors. A natural measure of the goodness of an estimator is its \emph{mean square error}, which coincides with its variance in the case of an unbiased estimator. By the independence of the realizations we have
\begin{equation}
\text{MSE}_{\widehat M_t}= \Var_\Pbb\big[\widehat M_t\big]=\frac{\Var_\Pbb \big[F[\psi,t]\big]}{R},
\end{equation}
\begin{equation}
\Var_\Pbb \big[F[\psi,t]\big]=\Ebb_\Pbb \left[ \left(F[\psi,t]- M_t\right)^2 \right] =\Ebb_\Pbb \left[ F[\psi,t]^2 \right] - {M_t}^2.
\end{equation}
Obviously, $\Var_\Pbb \big[F[\psi,t]\big]$ is a theoretical quantity and needs to be estimated; its natural unbiased estimator is the \emph{sample variance}. At the end, the natural unbiased estimator of the mean square error is
\begin{equation}
\widehat{\sigma}^{\,2}_t=\widehat{\text{MSE}}_{\widehat M_t}= \frac{1}{R(R-1)}\sum_{r=1}^{R}\left(F[\psi^r,t]-\widehat{M}_t\right)^2 =\frac 1 {R-1} \biggl( \frac 1 R \sum_r F[\psi^r,t]^2 - \widehat{M}_t^{\;2}\biggr).
\end{equation}
The quantity $\widehat{\sigma}_t$ is known as the \emph{sample standard error} of the estimate of the mean value $ M_t$.
If the realizations in the sample are statistically independent, as we have assumed, and $\Var_\Pbb \big[F[\psi,t]\big]$ is finite, the standard error $\widehat{\sigma}_t$ decreases with the square root of the sample size $R$:
\begin{equation}
\widehat{\sigma}_t\sim\frac{1}{\sqrt{R}}.
\end{equation}

Of particular  interest are the a posteriori quantum  expectation values of some selfadjoint operator $C$: $F[\psi, t] = \langle \psi(t)|C\psi(t)\rangle $. Note that to have these quantities for any $C$ in a basis in the space of the bounded selfadjoint operators is equivalent to give all the matrix elements of the a posteriori state $\rho(t)=|\psi(t)\rangle \langle \psi(t) |$. By Eqs.\ \eqref{eta(t)1} and \eqref{Mt} we get
\begin{equation}
M_t=\Ebb_\Pbb[\langle\psi(t)|C\psi(t)\rangle]= \Tr \left\{C\eta(t)\right\}.
\end{equation}
Now the estimator of $M_t$ takes the form
\begin{equation}
\widehat{M}_t=\frac{1}{R}\sum_{r=1}^{R}\langle\psi^r(t)|C\psi^r(t)\rangle,
\end{equation}
and the estimator of its mean square error becomes
\begin{equation}
\widehat{\sigma}^{\,2}_t=\frac{1}{R(R-1)}\sum_{r=1}^{R}\left(\langle\psi^r(t)|C\psi^r(t)\rangle-\widehat{M}_t\right)^2.
\end{equation}

Let us stress that the sample standard error $\widehat{\sigma}_t$ is a measure of the statistical fluctuations, not of the numerical errors in the simulations, such that the ones due to approximations or to the discretization of the time in solving the evolution equation.

\subsection{Homodyne photodetection}\label{sec:hp}
Let us consider as a first example the stochastic Schr\"{o}dinger equation corresponding to homodyne photodetection \cite{4} of the light emitted by a two-level atom stimulated by a perfectly coherent laser in resonance with the atomic frequency \cite[Sections 8.1.3.2 and 9.2]{3}. We consider the ideal case in which all the emitted light is detected and no other dissipative contribution is present, apart from the emission of light.

Let $|1\rangle$  ($|0\rangle$) be the excited (ground) state and let $\sigma_x$, $\sigma_y$, $\sigma_z$ be the usual Pauli matrices and $\sigma_{-}$ and $\sigma_{+}$ be the lowering and rising operators; then, $\sigma_++\sigma_-=\sigma_x$, \ $\ri\left(\sigma_-+\sigma_+\right)=\sigma_y$ and $\sigma_+\sigma_-$ is the projection on the excited state.

The model we are considering is determined by the SSE \eqref{SSE1}, \eqref{mj} with $d=1$,
\begin{subequations}\label{model1}
\begin{equation}
H(t)= \frac {\omega_0}2\, \sigma_z - \frac{\Omega_R}2\left( \re^{\ri \omega_0t}\sigma_- +\re^{-\ri \omega_0t}\sigma_+\right), \qquad \omega_0>0, \quad \Omega_R\geq 0,
\end{equation}
\begin{equation}
R(t)= \sqrt{\gamma}\, \re^{\ri \left(\omega_0t+\theta\right)}\sigma_-\, , \qquad \gamma>0.
\end{equation}
\end{subequations}
In this model the frequencies of the atom, of the stimulating laser and of the local oscillator are equal and given by $\omega_0$; $\Omega_R$ is the \emph{Rabi frequency} ($\Omega_R^{\;2}$ is proportional to the laser intensity), $\gamma$ is the \emph{natural linewidth} of the atom ($1/\gamma$ is the relaxation time) and $\theta$ is the phase shift of the local oscillator with respect to the emitted light. Homodyne detection is sensitive to $\theta$, as discussed in Section \ref{sec:nlSSE}; here, we take $\theta=\pi/2$.

The explicit time dependencies can be eliminated by a unitary transformation:
\begin{equation}
\check \psi(t):= \exp\left\{\frac \ri 2 \, \omega_0 \sigma_z t\right\} \psi(t).
\end{equation}
Then, by Eqs.\ \eqref{mj}, \eqref{SSE1}, \eqref{model1} we get the SSE in the rotating frame:
\begin{multline}\label{3.2.1}
\rd\check\psi(t)=-\ri H_L\check \psi(t)\,\rd t+ \frac{\gamma}{2}\left(m_y(t)\ri \sigma_{-}- \sigma_{+}\sigma_{-}- \frac{1}{4}\,m_y(t)^2 \right)\check \psi(t)\,\rd t\\
{}+\sqrt{\gamma}\left(\ri\sigma_{-}-\frac{1}{2}\,m_y(t) \right)\check\psi(t)\,\rd W(t),
\end{multline}
\begin{equation}
H_L=-\frac{\Omega_R}{2}\,\sigma_x, \qquad m_y(t)=\langle \check \psi(t)|\sigma_y \check \psi(t) \rangle.
\end{equation}
Moreover, by Eqs.\ \eqref{Gir} and \eqref{mj}, the cumulated output (the integrated \emph{homodyne photocurrent}) is given by
\begin{equation}\label{mod1_out}
B(t)=W(t)+ \sqrt{\gamma}\int_0^t m_y(s)\,\rd s.
\end{equation}

The master equation corresponding to the SSE \eqref{3.2.1} is
\begin{equation}
\frac{\rd\check \eta(t)}{\rd t}=\check \Lcal\big[\check \eta(t)\big], \qquad \check \Lcal[\varrho]= \frac{\ri \Omega_R}{2}[\sigma_x,\varrho] +\gamma \sigma_-\varrho\sigma_+ -\frac\gamma 2 \left\{\sigma_+\sigma_-,\varrho\right\}.
\end{equation}
This equation can be easily solved \cite{4} and we get, with the initial condition $\eta(0)=|0\rangle \langle 0|$ and $\Omega_R^{\,2}>\gamma^2/16$, \cite[Section 8.2.2.2]{3}
\begin{equation}\label{1eta1}
\eta(t)_{11}=\langle 1|\eta(t)|1\rangle = v_+ \re^{-a_+ t}+v_-\re^{-a_- t} + \frac {\Omega_R^{\,2}}{2\Omega_R^{\,2}+\gamma^2}, \qquad \Tr\{\sigma_x \eta(t)\}=0,
\end{equation}
\begin{equation}\label{y}
\Tr\{\sigma_y \eta(t)\}=u_+\re^{-a_+ t}+u_-\re^{-a_- t} - \frac {\Omega_R\gamma}{\Omega_R^{\,2}+\gamma^2/2},
\end{equation}
\[
u_\pm=\frac{\Omega_R
 \left[ \gamma \sqrt{\Omega_R^{\,2}-\gamma^2/16}\mp \ri \left(\Omega_R^{\,2}-\gamma^2/4\right)\right]}{\sqrt{\Omega_R^{\,2}-\gamma^2/16}\left(2\Omega_R^{\,2}+\gamma^2\right)},
 \]
\[
v_\pm=\frac{\Omega_R^{\,2}\left(\mp 3\ri \gamma /4 -
 \sqrt{\Omega_R^{\,2}-\gamma^2/16}\right)}{2\sqrt{\Omega_R^{\,2}-\gamma^2/16}\left(2\Omega_R^{\,2}+\gamma^2\right)}
, \qquad
a_\pm= \frac 3 4 \, \gamma \pm \ri \sqrt{\Omega_R^{\,2}-\frac{\gamma^2}{16}}.
\]
Note that
\begin{equation}\label{meanOUT}
\Ebb_\Pbb[B(t)]=\sqrt{\gamma}\int^t_0 \Tr\{\sigma_y \eta(s)\}\rd s.
\end{equation}

To simulate this model we use the Euler algorithm to get an approximation for the state vector $\check\psi$, with a correction to maintain the normalization. We discretize the time and set $t_n= n \Delta t$; then, the algorithm  takes the form
\begin{subequations}\label{recursion}
\begin{equation}\label{recur}
\psi_{n+1}=\check\psi_n+A_1(\check\psi_n)\Delta t+A_2(\check\psi_n)\Delta W_n,
\end{equation}
\begin{equation}
\check\psi_{n+1}=\frac{\psi_{n+1}}{\norm{\psi_{n+1}}},
\end{equation}
where $\Delta W_n= W(t_{n+1})-W(t_n)=Z_n \sqrt{\Delta t}$, $Z_0,\ldots,Z_n,\ldots$ is a sequence of independent random variables with standard normal distribution, and the functions $A_1, A_2$ are given by
\begin{gather}
A_1(\psi)=-\ri H_L \psi+ \frac{\gamma}{2}\left(\langle\psi|\sigma_y\psi\rangle \ri\sigma_{-}- \sigma_{+}\sigma_{-}- \frac{1}{4}\,\langle\psi|\sigma_y\psi\rangle^2 \right) \psi,\\
A_2(\psi)=\sqrt{\gamma}\left(\ri\sigma_{-}-\frac{1}{2}\langle\psi|\sigma_y \psi\rangle\right)\psi.
\end{gather}
As initial condition we take the ground state
\begin{equation}
\check \psi_0 =\psi_0 =|0\rangle,
\end{equation}
\end{subequations}
\begin{figure}[h]
\includegraphics[scale=0.7]{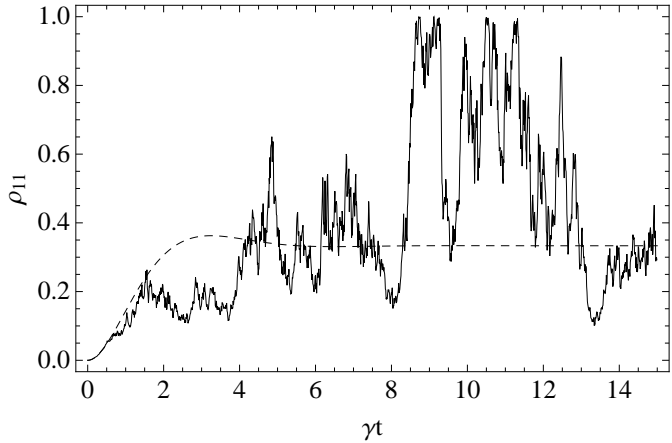}
\caption{A single realization of the occupation of the excited state $\rho_{11}=\left|\langle1|\check\psi(t)\rangle\right|^2$ computed from Eqs.\  \eqref{recursion} for the parameters: $\Omega_R=1$, $\gamma=1$, $\Delta t=0.01$. The dashed line is the plot of the component $\eta(t)_{11}$ of the exact solution Eq. (\ref{1eta1}).}\label{mod}
\end{figure}
By construction, $\psi_n$ is an approximation of $\check \psi(t_n)$, so that $
\psi(t_n)\simeq \exp\left\{-\frac \ri 2 \, \omega_0 t_n \sigma_z\right\} \psi_n$. Correspondingly, by \eqref{mod1_out}, the approximation of the integrated homodyne current is
\begin{equation}\label{approxOUT}
B(t_n)\simeq B_n= \sum_{k=0}^{n-1}\left( \Delta W_k + \sqrt{\gamma} \langle \psi_k|\sigma_y\psi_k\rangle \Delta t\right).
\end{equation}
Let us note that, by the properties of the Wiener process, $\Delta W_n/\sqrt{\Delta t}$, $n=1, 2, \ldots$, is a sequence of independent and identically distributed random variables with standard normal distribution.

%
%
%
%
%
%
%
%
%

\begin{figure}[h]
\includegraphics[scale=0.7]{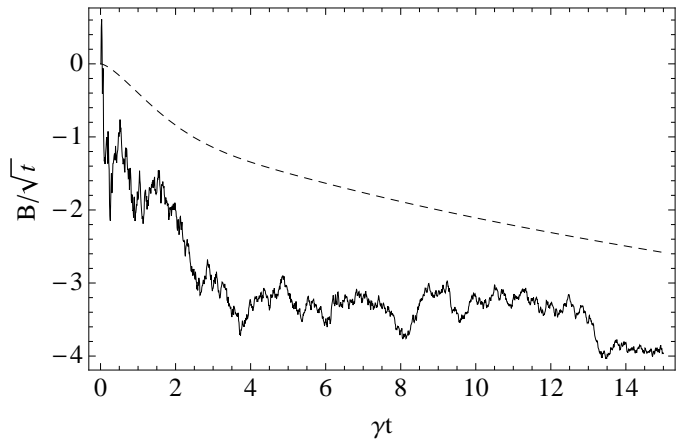}
\caption{A single realization of the output $B(t)/\sqrt{t}$ computed from Eq.\  \eqref{approxOUT} and the plot of the mean output from Eqs.\ \eqref{meanOUT} and \eqref{y} for the parameters: $\Omega_R=1$, $\gamma=1$, $\Delta t=0.01$. }\label{modX}
\end{figure}

\begin{figure}[h]
\includegraphics[scale=0.7]{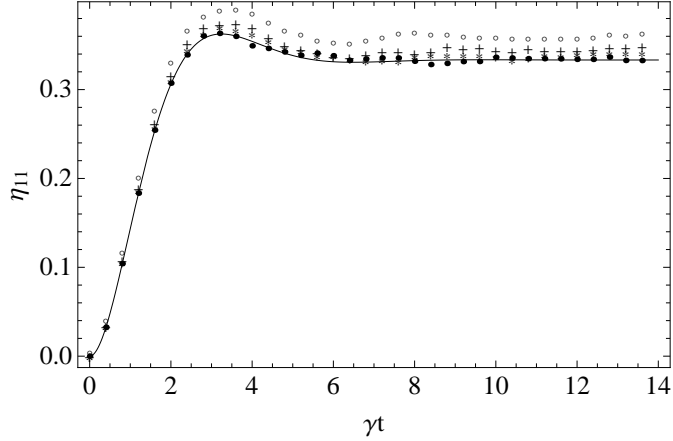}
\caption{The average over 10000 realizations of the homodyne photodetection (\ref{3.2.1}) for the driven two-level atom. The realizations are computed with the Euler algorithm for the parameters $\Omega=1$, $\gamma=1$. The dots show $\eta_{11}$ computed from the average over the realizations for the different sizes of time steps $\Delta t_1=0.01$ (dots), $\Delta t_2=0.05$ (stars), $\Delta t_3=0.1$ (pluses) and $\Delta t_4=0.2$ (circles). The solid line represents the analytical solution for $\eta_{11}$ according to \eqref{1eta1}. The statistical errors have the same size as the graphic symbols used in the figure.}\label{mod1}
\end{figure}

The results of the simulation are shown in Figs.\ \ref{mod}--\ref{mod1}. A single realization is shown in Fig.\ref{mod} for the occupation of the excited state.
In Fig.\ \ref{modX} we plot a single realization of the output and, for comparison, its mean.
Finally, in Fig.\ \ref{mod1} we analyse the dependence of the simulation algorithm on the time step size. It is clearly seen that the quality of the simulation with the help of Euler algorithm decreases with increasing time step. In principle, extrapolation techniques can correct the results. However, it is more efficient to use the higher order scheme such as the Platen scheme as we shall demonstrate in the next section.

\subsection{Damped harmonic oscillator}
Another typical example of an open system in the Markovian regime is the stochastic Schr\"{o}dinger equation \eqref{SSE1} for the damped harmonic oscillator \cite[Section 7.3.1.2]{4}:
\begin{multline}\label{3.1.1}
\rd\psi(t)=\frac{\gamma}{2}\left(\langle a+a^\dag\rangle_{\psi(t)}a-a^\dag a-\frac{1}{4}\langle a+a^\dag\rangle^{\;2}_{\psi(t)}\right)\psi(t)\,\rd t\\
{}+\sqrt{\gamma}\left(a-\frac{1}{2}\langle a+a^\dag\rangle_{\psi(t)}\right)\psi(t)\,\rd W(t),
\end{multline}
\[
\langle a+a^\dag\rangle_{\psi}=\big\langle \psi\big|\left(a+a^\dag\right)\psi\big\rangle.
\]
The SSE \eqref{3.1.1} could be obtained as Eq.\ \eqref{3.2.1} by considering an harmonic oscillator with homodyning and by performing a unitary transformation. However, here the interest in this model is mainly to use it for introducing a higher order numerical scheme.

\begin{figure}[h]
  \includegraphics[scale=0.7]{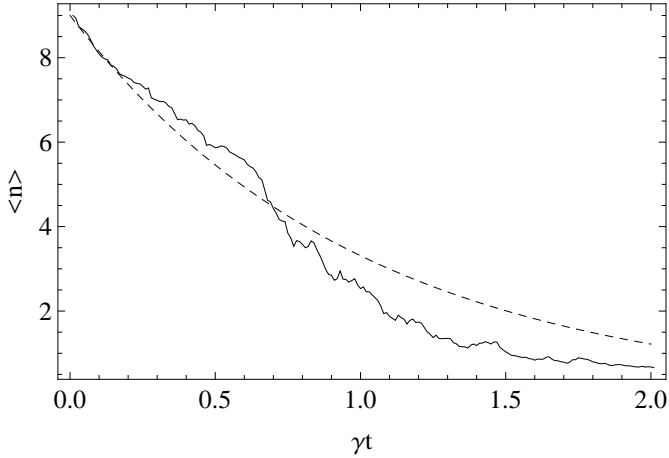}
  \caption{A single realization of the damped harmonic oscillator for the a posteriori expectation of $n=a^\dag a$ computed from Eq.\ (\ref{3.1.1}) with parameters: $\gamma=1$ and $\Delta t=0.01$. The initial condition is $|n_0=9\rangle$. The dashed line shows the exact solution for the master equation associated to the SSE \eqref{3.1.1} according to \cite{4}.}\label{mod3}
  \end{figure}

\begin{figure}[h]
  \includegraphics[scale=0.7]{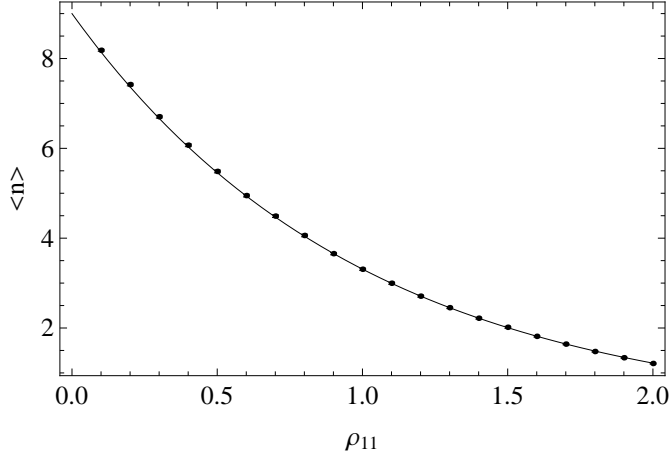}
  \caption{The average over 1000 realizations of the damped harmonic oscillator by Eq.\ (\ref{3.1.1}). The dots show $<n>$ (the mean number of photons) computed from the average over the realizations. The continuous line represents the analytical solution for $<n>$ according to reference \cite{4}. The simulation was performed for the following parameters: $\gamma=1$ and $\Delta t=0.01$ with the initial condition $|n_0=9\rangle$. The statistical errors have the same size as the dots.}\label{mod11}
  \end{figure}

As an example, the initial state is taken to be $\psi_0=|n_0=9\rangle$ (a pure Fock state with 9 photons) and the Hilbert space has been truncated at $n_{\mathrm{max}}=12$ which means that the simulation was performed in a subspace of dimension $N=13$. The size of the time steps is $\Delta t = 0.02$.
To simulate this model we use the second-order weak scheme of Platen \cite{4}. This algorithm has the form
\begin{eqnarray}
\psi_{n+1}&=&\psi_n+\frac{1}{2}\left(D_1(\tilde{\psi}_n)+D_1(\psi_n)\right)\Delta t\nonumber\\
&&{}\ {}+\frac{1}{4}\left(D_2(\psi^{+}_n)+D_2(\psi^{-}_n)+2D_2(\psi_n)\right)\Delta W_n\nonumber\\
&&{}\ {}+\frac{1}{4}\left(D_2(\psi^{+}_n)-D_2(\psi^{-}_n\right)\{(\Delta W_n)^2-\Delta t\}\Delta t^{-1/2}\nonumber,
\end{eqnarray}
where
\begin{eqnarray}
\tilde{\psi}_n&=&\psi_n+D_1(\psi_n)\Delta t+D_2(\psi_n)\Delta W_n,\nonumber\\
\psi^{\pm}_n&=&\psi_n+D_1(\psi_n)\Delta t\pm D_2(\psi_n)\sqrt{\Delta t}.\nonumber
\end{eqnarray}
For the model under consideration the functions $D_1$ and $D_2$ are
\begin{eqnarray}
D_1(\psi)&=&\frac{\gamma}{2}\left(\langle a+a^\dag\rangle_{\psi}a-a^\dag a-\frac{1}{4}\langle a+a^\dag\rangle^2_{\psi}\right)\psi,\nonumber\\
D_2(\psi_t)&=&\sqrt{\gamma}\left(a-\frac{1}{2}\langle a+a^\dag\rangle_{\psi}\right)\psi.\nonumber
\end{eqnarray}
A single realization for the damped harmonic oscillator is shown in Fig.\ \ref{mod3}.
The number of photons, computed from the average of 1000 realizations is shown in Fig.\ \ref{mod11}.

\section{SSEs with memory effects}\label{sec:memory}
One of the methods for the introduction of memory effects in the system is to start from the lSSE \eqref{2.1.3}, but with random coefficients $H(t)$, $R_j(t)$ and with the white noise replaced by some coloured noise. In this way we get memory in the dynamical equations, while complete positivity of the dynamical maps and the continuous measurement interpretation are preserved \cite{1,2,apr}.

In this section we want to consider a very particular case of non-Markovian SSE and to use it to illustrate two methods of numerical approximations: the simulation of the SSE and an approximation derived in \cite{apr} based on the Nakajima-Zwanzig projection method.
Specifically, we start with a lSSE driven by a coloured noise with non-random coefficients. In this way the memory is encoded in the driving noise of the lSSE, not in the coefficients. In this case, the new lSSE will be norm-preserving, as in Section \ref{sec:randomunitary}, and will represent a quantum system evolving under a random Hamiltonian dynamics, while the Hamiltonian is very singular and produces dissipation.

\subsection{Coloured noise}\label{sec:colnoise}
Let us consider a one-dimensional driving noise $X(t)$ and two non-random operators $C$ and $D$ on $\Hscr$. The starting point is the basic linear stochastic Schr\"{o}dinger equation
\begin{equation}\label{4.15}
\rd\psi(t)=C\psi(t)\,\rd t+D\psi(t)\, \rd X(t).
\end{equation}
The simplest choice of a coloured noise is the stationary Ornstein-Uhlenbeck (O-U) process defined by
\begin{equation}\label{O-U}
X(t)=\re^{-k t}\,\frac{Z}{\sqrt{2k}}+\int_{0}^{t}\re^{-k(t-s)}\, \rd W(s), \qquad k>0,
\end{equation}
where $W(t)$ is a one-dimensional Wiener process, defined on the stochastic basis $\big(\Omega, \Fscr, (\Fscr_t)$, $\mathbb{P}\big)$, and $Z$ is a standard normal random variable (mean $0$ and variance $1$); $Z$ is $\Fscr_0$-measurable, which means that it is independent from the Wiener process. The O-U process $X(t)$ is a Gaussian process with zero mean and correlation function
\begin{equation}
\Ebb_\Pbb[X(t)X(s)]=\frac{\re^{-k|t-s|}}{2k}.
\end{equation}
It satisfies the stochastic differential equation
\begin{equation}\label{4.1.18}
\rd X(t)=-k X(t)\,\rd t+\rd W(t),\qquad X(0)=Z/\sqrt{2k}.
\end{equation}
Formally, Eq.\ (\ref{4.15}) is driven by the derivative of the O-U process, whose two-time correlation is no more a $\delta$-function, as in the case of white noise, but it is formally given by
\begin{equation}
\Ebb_\Pbb[\dot X(t)\dot X(s)]=\delta (t-s)-\frac{k}{2}\,\re^{-k|t-s|}.\nonumber
\end{equation}
Note that the Markovian regime is recovered in the limit $k\downarrow 0$. It is then straightforward that Eq.\ (\ref{4.15}) can be rewritten in the form
\begin{equation}
\rd\psi(t)=\bigl(C-kX(t)D\bigr)\psi(t)\, \rd t+D\psi(t) \rd W(t);
\end{equation}
the initial condition is a wave function $\psi_0\in \Hscr$, such that $\Arrowvert\psi_0\Arrowvert^2=1$.

As discussed in Section \ref{sec:lSSE} for the Markovian case, to construct consistent probabilities we need the process $\Arrowvert\psi(t)\Arrowvert^2$ to be a martingale. By It\^{o} calculus rules (see Appendix \ref{Calc}), the stochastic differential of $\Arrowvert\psi(t)\Arrowvert^2$ turns out to be
\begin{multline}\label{agreement}
\rd\langle\psi(t)|\psi(t)\rangle=\langle \rd\psi(t)|\psi(t)\rangle+\langle \rd\psi(t)|\rd\psi(t)\rangle+ \langle \psi(t)|\rd\psi(t)\rangle\\
{}=\langle\psi(t)|\left[C^{\dag}+C-k X(t)(D^{\dag}+D)+D^{\dag}D\right]\psi(t)\rangle\rd t
+ \langle\psi(t)|(D^{\dag}+D)\psi(t)\rangle\rd W(t).
\end{multline}
Then, the process $\Arrowvert\psi(t)\Arrowvert^2$ can be a martingale only if the term in front of $\rd t$ is equal to zero.  For it we must have
\begin{equation*}
C^{\dag}+C+D^{\dag}D=kX(t)(D^{\dag}+D),\qquad \forall t,
\end{equation*}
which implies $D^\dag+D=0$ and $C^{\dag}+C+D^{\dag}D=0$. These conditions impose that there are two self-adjoint operators $L$ and $H_0$ such that $D = -\ri L$ and $C = -\ri H_0 - \frac{1}{2}\,L^2$. As a consequence the initial Eq.\ (\ref{4.15}) becomes
\begin{equation}\label{4.1.8}
\rd\psi(t)=-\ri\left[\bigl(H_0-k X(t)L\bigr)\rd t+ L \,\rd W(t)\right]\psi(t) -\frac{1}{2}\,L^2\psi(t) \, \rd t.
\end{equation}
Apart the further randomness introduced by the term with $X(t)$, we are in the same situation of Eq.\ \eqref{Uevol} and the evolution of the quantum system is then completely determined by the time-dependent, random Hamiltonian
\begin{equation}
H(t)=H_0+\bigl(\dot W(t)-k X(t)\bigr)L.
\end{equation}
Let us stress that it is a formal expression, due to the presence of the white noise $\dot W(t)$.

As in Section \ref{sec:randomunitary} the model we have constructed represents a dissipative evolution, now with memory, but without observation of the quantum system. In this case there is no change of probability, $\Pbb$ is also the physical probability, and $\norm{\psi(t)}=1$, $\forall t$. The theory can be generalized \cite{2} by taking the operators $C$ and $D$ dependent on the O-U process; in this way also a true continuous observation can be introduced.

Let us stress that the class of models presented in this section is very peculiar. The process $\big(X(t),\, \psi(t)\big)_{t\geq 0}$ satisfies the couple of SDEs \eqref{4.1.18} and \eqref{4.1.8}, whose coefficients depend only on the process of time $t$; then, this composed process is Markovian.

\subsection{Projection techniques and closed master equations with memory}
As in the Markovian case, the average statistical operator (the a priori state) can be introduced:
\begin{equation}\label{aprioriM}
\eta(t)=\Ebb_\Pbb[|\psi(t)\rangle\langle \psi(t)|].
\end{equation}
However, to get a closed equation for $\eta(t)$ is not a trivial task \cite{apr}; the final result is a generalized master equation with memory. The important point is that the complete positivity of the map $\eta(0)\mapsto \eta(t)$ is guaranteed by the stochastic representation \eqref{aprioriM}. We illustrate these techniques on the model of Section \ref{sec:colnoise}.

Let us define the process
\begin{equation}
\rho(t)=|\psi(t)\rangle\langle\psi(t)|.
\end{equation}
By Eq.\ \eqref{4.1.8} and It\^{o} rules, we can compute the stochastic differential of $\rho(t)$; the result is the stochastic master equation
\begin{gather}\label{eq}
\rd\rho(t)=\mathcal{L}(t)[\rho_t]\,\rd t+\Rcal[\rho_t]\,\rd W(t)\equiv \Lcal_0[\rho_t]\,\rd t+\Rcal[\rho_t]\,\rd X(t),
\\
\mathcal{L}(t)=\Lcal_0-kX(t)\Rcal, \qquad \mathcal{L}_0[\rho]=-\ri[H_0, \rho]-\frac{1}{2}\bigl[L,\left[L,\rho\right]\bigr], \qquad \Rcal[\rho]=-\ri [L,\rho].
\end{gather}

By taking the mean of Eq.\ \eqref{eq} and by recalling that $W$ has mean zero and increments independent from the past we get
\begin{equation}\label{doteta}
\dot \eta(t)=\Lcal_0[\eta(t)]-k \Rcal\bigl[\Ebb_\mathbb{P}[X(t)\rho(t)]\bigr],
\end{equation}
which is a kind of master equation with  non-Markovian effects introduced by the last term. However, this master equation is not closed, because the $X(t)$ and $\rho(t)$ are random and not independent.

A closed equation can be obtained by using the Nakajima-Zwanzig method and the generalized master equation one obtains in this way can be the starting point for some approximations \cite{apr}.
Indeed, the operation of taking the mean is a projection in the space of random trace class operators. We can think to $\eta(t)$ as the \emph{relevant} part of $\rho(t)$, while $\rho_\bot(t)=\rho(t)-\eta(t)$ is the \emph{non relevant} part. As we took a non-random initial state, we have $\rho(0)=\eta(0)$, $\rho_\bot(0)=0$. By taking the stochastic differential of $\rho_\bot(t)$ and by using Eqs.\ \eqref{eq} and \eqref{doteta}, we get
the system of equations
\begin{subequations}\label{NZ}
\begin{equation}\label{NZ1}
\dot \eta(t)=\Lcal_0[\eta(t)]-k \Rcal\bigl[\Ebb_\Pbb[X(t)\rho_\bot(t)]\bigr],
\end{equation}
\begin{multline} \label{NZ2}
\rd \rho_\bot(t)=\Lcal_0\left[\rho_\bot(t)\right]\rd t - k \Rcal\bigl[X(t)\rho_\bot(t)-\Ebb_\Pbb[X(t)\rho_\bot(t)]\bigr]\rd t \\ {}+\Rcal\left[ \rho_\bot(t)\right]\rd W(t)+\Rcal\left[\eta(t)\right]\rd X(t).
\end{multline}
\end{subequations}

Let us introduce now the propagator of the homogeneous part of Eq.\ \eqref{NZ2}, which is defined by the SDE
\begin{equation}\label{propag}
\Vcal(t,s)=\openone+\int_{s}^{t}\bigl(\Lcal_0- k\Rcal\circ \Xcal(r) \bigr)\circ\Vcal(r,s)\,\rd r +  \int_{s}^{t}\Rcal \circ \Vcal(r,s) \, \rd W(r),
\end{equation}
where $\circ$ denotes the composition of maps and $\Xcal(t)$ is the map $\rho\mapsto X(t)\rho-\Ebb_\Pbb[X(t)\rho]$. Then, the formal solution of the Eq.\ \eqref{NZ2} with $\rho_\bot(0)=0$ can be written as
\begin{equation}\label{rhobot}
\rho_\bot(t)=-k\int_0^t \Vcal(t,s)\circ\Rcal[X(s)\eta(s)]\,\rd s
+ \Vcal(t,0)\circ\int_0^t \Vcal(s,0)^{-1}\circ\Rcal[\eta(s)]\,\rd W(s).
\end{equation}
In the last term we used $\Vcal(t,0)\circ\Vcal(s,0)^{-1}$ instead of $\Vcal(t,s)$ in order to have an adapted integrand in the stochastic integral, as required by the It\^o formulation. By inserting the expression \eqref{rhobot} into Eq.\  \eqref{NZ1} we get the generalized master equation for the a priori states
\begin{multline}\label{GME}
\dot{\eta}_t=\Lcal_0[\eta(t)]+k^2\int_{0}^{t}\Rcal\circ \Ebb_\Pbb[X(t)X(s)\Vcal(t,s)]\circ \Rcal[\eta(s)]\,\rd s\\
{}-k\Ebb_{\mathbb{P}}\biggl[X(t)\Rcal\circ \Vcal(t,0) \circ \int_0^t \Vcal(s,0)^{-1}\circ\Rcal[\eta(s)]\,\rd W(s)\biggr] .
\end{multline}

Equation \eqref{GME} is very complicated, but it is useful as a starting point to find approximations. In \cite{apr} it is suggested to take the non random approximation of the propagator \eqref{propag}: $\Vcal(t,s) \simeq \re^{\Lcal_0(t-s)}$. Then, the mean values in \eqref{GME} can be computed and the generalized master equation takes the form
\begin{equation}\label{48}
\dot\eta(t)\simeq \Lcal_0[\eta(t)]
+\frac{k}{2}\int_{0}^{t}\left[L,\re^{(\mathcal{L}_0-k) (t-s)}\bigl[\left[L,\eta(s)\right]\bigr]\right]\rd s.
\end{equation}

\subsection{A non Markovian model: a dissipative qubit}

In this section we introduce a very simple example based on a qubit with dissipation in order to have a toy model with a non Markovian dynamics for which we can do stochastic simulations and test the approximation \eqref{48}.

Let us take a two-level system as in Section \ref{sec:hp} and consider the stochastic dynamics \eqref{4.1.8} with
\begin{equation}
H_0=\frac{\omega_0}2\,\sigma_z, \quad \omega_0>0, \qquad L=\sqrt{\frac\gamma 2}\,\sigma_y, \quad \gamma>0.
\end{equation}
Then, the SSE \eqref{4.1.8} becomes
\begin{subequations}\label{2SSE}
\begin{gather}
\rd \psi_1(t) = -\frac 1 2 \left( \frac \gamma 2 +\ri \omega_0\right) \psi_1(t) \,\rd t -\sqrt{\frac \gamma 2}\,\psi_2(t)\,\rd X(t),
\\
\rd \psi_2(t) = -\frac 1 2 \left( \frac \gamma 2 -\ri \omega_0\right) \psi_2(t) \,\rd t +\sqrt{\frac \gamma 2}\,\psi_1(t)\,\rd X(t).
\end{gather}
\end{subequations}
The O-U process $X(t)$ is given by Eq.\ \eqref{O-U} and its stochastic differential by \eqref{4.1.18}.

For this model we have
\begin{equation}
\mathcal{L}_0[\rho]=-\frac{\ri\omega_0}2\left[\sigma_z, \rho\right]
-\frac{\gamma}{4}\bigl[\sigma_y,\left[\sigma_y,\rho\right]\bigr].
\end{equation}
By representing the states in the Bloch sphere, the master equation $\dot\xi(t)=\Lcal_0[\xi(t)]$ can be explicitly solved and the right hand side of Eq.\ \eqref{48} can be given an explicit expression. Indeed, by writing
\begin{equation}\label{Bloch}
\eta(t)=\frac 1 2 \left[ \openone + \vec x(t) \cdot \vec \sigma\right],
\end{equation}
from Eq.\ \eqref{48} we get
\begin{equation}\label{Blocheqs}
\begin{cases}\displaystyle
\dot x(t)=-\omega_0y(t) -\gamma x(t) +k\gamma \int_0^t \re^{-(k+\gamma)(t-s)}x(s)\,\rd s,
\\
\dot y(t)=\omega_0x(t),
\\ \displaystyle
\dot z(t)=-\gamma z(t) +k\gamma\int_0^t \re^{-(k+\gamma/2)(t-s)}\left(\cos \nu (t-s) - \frac\gamma{2\nu}\,\sin \nu(t-s)\right) z(s)\,\rd s.
\end{cases}
\end{equation}
We assume to have $\omega_0>\gamma/2$ and set $\nu=\sqrt{\omega_0^{\;2}- \gamma^2/4}$. Recall that \eqref{Bloch} and \eqref{Blocheqs} give an approximation of the a priori states.

Equations \eqref{Blocheqs} can be solved by Laplace transform techniques or, equivalently, by increasing the degrees of freedom. Let us set
\begin{subequations}
\begin{gather}
\xi(t)=\gamma\int_0^t \re^{-(k+\gamma/2)(t-s)}\cos \bigl(\nu (t-s)\bigr)  z(s)\,\rd s,
\\
\epsilon(t)=- \frac{\gamma^2}{2\nu}\int_0^t \re^{-(k+\gamma/2)(t-s)}\sin \bigl(\nu(t-s)\bigr) z(s)\,\rd s,
\\
\zeta(t)=\gamma\int_0^t \re^{-(k+\gamma)(t-s)}x(s)\,\rd s.
\end{gather}
\end{subequations}
Then, Eqs.\ \eqref{Blocheqs}  reduce to the two decoupled systems of linear equations with constant coefficients
\begin{subequations}\label{approxA}
\begin{equation}\label{part1}
\begin{cases}
\dot x(t)=-\omega_0y(t) -\gamma x(t) +k\zeta(t),
\\
\dot y(t)=\omega_0x(t),
\\ \displaystyle
\dot \zeta(t)=-(k+\gamma)\zeta(t)+\gamma x(t),
\end{cases}
\end{equation}
\begin{equation}\label{part2}
\begin{cases}
\dot \xi(t)=-\left(k+\frac \gamma 2 \right)\xi(t) +\frac{2\nu^2}\gamma\, \epsilon(t)+\gamma z(t),
\\
\dot \epsilon(t)=-\left(k+\frac \gamma 2 \right)\epsilon(t)-\frac \gamma 2 \,\xi(t),
\\ \displaystyle
\dot z(t)=-\gamma z +k\bigl(\xi(t)+\epsilon(t)\bigr).
\end{cases}
\end{equation}
\end{subequations}

To get the mean state $\eta(t)$ we can now use stochastic simulations or the analytical approximation of Eqs.\ \eqref{Bloch}, \eqref{approxA}. We concentrate on the study of the occupation of the excited state $\eta(t)_{11}=\frac 1 2 \left[1+z(t)\right]$. Let us stress that it is easy to prove that $\lim_{t\to +\infty}\eta(t)_{11}=0.5$.

\begin{figure}[t]
  \includegraphics[scale=0.7]{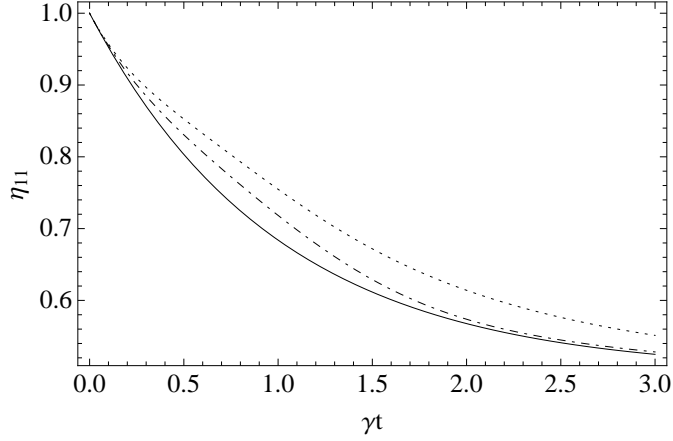}
  \caption{Plot of the mean occupation of the excited state $\eta_{11}$ computed from Eqs.\  \eqref{part2} for the parameters: $\gamma=1$, $\nu=3$ and $k=0$ (solid line), $k=1$ (dot-dashed line), $k=2$ (dotted line).}\label{Q1}
  \end{figure}
Let us start from the analytical solution.
In Figure \ref{Q1} we plot $\eta(t)_{11}$ obtained by solving system \eqref{part2} by using the internal function of Mathematica ``DSolve [ ]''. The choice of parameters is $\gamma=1$, $ \omega_0=\sqrt{37}/2$ (which gives $\nu=3$) and $k=0,\; 1, \;2$; recall that $k=0$ is the Markovian case. The initial state is  $\eta(0)=\begin{pmatrix} 1 & 0 \\ 0&0\end{pmatrix}$.

We can say that in this model the effect of memory (increasing $k$) is to modify and to slow down the decay.

\begin{figure}[h]
  \includegraphics[scale=0.7]{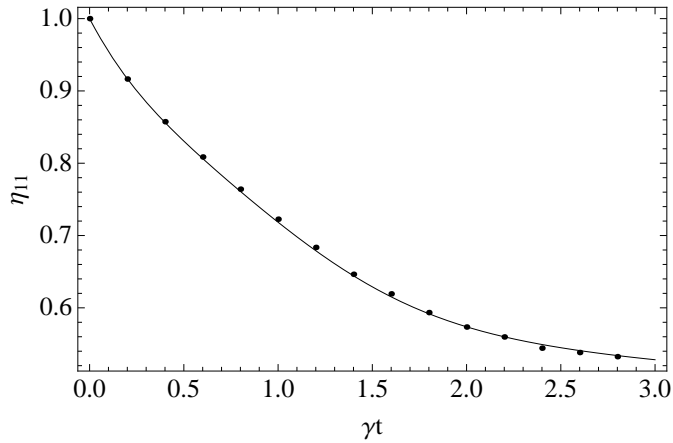}
  \caption{Plot of the mean occupation number of the excited state for the parameters $\gamma=1$, $ \omega_0=\sqrt{37}/2$, $k=1$, $\Delta t=0.01$. The solid line comes from the analytical approximation, while the dots from the stochastic simulations.}\label{Q2}
  \end{figure}

\begin{figure}[h]
 \includegraphics[scale=0.7]{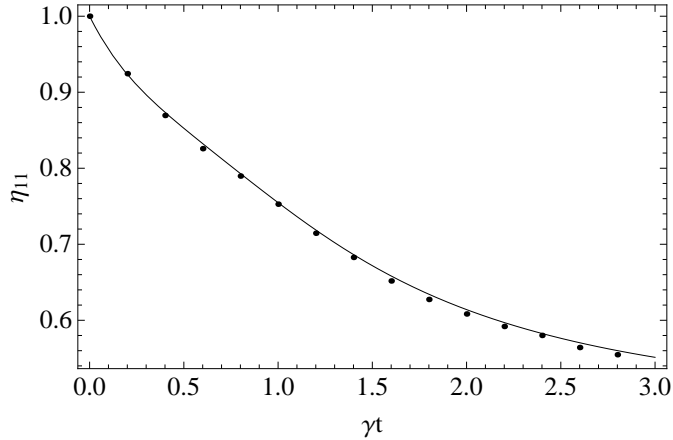}
  \caption{Plot of the mean occupation number of the excited state for the parameters $\gamma=1$, $ \omega_0=\sqrt{37}/2$, $k=2$, $\Delta t=1/200$. The solid line comes from the analytical approximation, while the dots from the stochastic simulations.}\label{Q3}
  \end{figure}

However Eqs.\ \eqref{part2} are approximated, but we can compare this solution with the simulations based on the exact equations \eqref{2SSE}, \eqref{4.1.18}. We use the Euler algorithm applied to the Markov process $\big(X(t),\psi_1(t),\psi_2(t)\big)_{t\geq 0}$ with normalization of $\psi(t)$ at every step as in Section \ref{sec:hp} (10000 realizations). In Figures \ref{Q2} and \ref{Q3} the dots comes from the simulations and the solid line from the analytical approximation; we see an extremely good agreement of simulations and approximated analytical solution.

\section{Conclusions}\label{sec:concl}

The theory of linear and nonlinear SSEs has been presented in the Markovian diffusive case. Moreover we have discussed their links with the dissipative dynamics of open systems and with measurements in continuous time. Two simple cases have been used to show how to make stochastic simulations based on the SSE. A two-level atom with homodyne detection has been used to show the Euler algorithm, while the Platen algorithm was illustrated in the case of a damped harmonic oscillator.

Then, we have shown how to use coloured noise in order to construct non Markovian models. Now the average state does not satisfy the usual Markovian quantum master equation. However, by adapting the Nakajima-Zwanzig projection method, it is possible to arrive to a generalized master equation and we have shown how to get an approximate solution for this equation. On the other side, the original SDEs can be simulated and the exact solution can be obtained up to numerical errors and statistical fluctuations. 

In a concrete model of a dissipative qubit we have compared the analytical approximation with the stochastic simulation of the exact equation. Such a comparison gives a strong support to the proposed approximation. This gives confidence in the possibility of studying more elaborated Markovian models, for which the two computational ways of treating them are open: analytic approximations and stochastic simulations. In the proposed model we see also some effects of the non-Markovianity: there is a slowdown of the decay and and a modification of its functional form.

\section{ACKNOWLEDGEMENTS}
This work is based upon research supported by the South African Research Chair Initiative of the Department of Science and Technology and national Research Foundation. 
\appendix
\section{Appendix: some theory of random processes}\label{appen}
\subsection{The Wiener Process}\label{Wiener}
A \emph{standard Wiener process} $\{W(t)\}_{t\geq 0}$  is a continuous Gaussian process starting from 0, with independent and stationary increments,
with mean zero and variance proportional to $t$; in particular, $
\mathbb{E}[W(t)]=0$ and $ \Cov[W(t)W(s)]=\mathbb{E}[W(t)W(s)]=\min(t,s)$.

Due to the Gaussianity and the independence of the increments, if we take a sequence of times $0\leq t_0<t_1<\cdots<t_n$ and set $Z_k=\frac{W(t_k)-W(t_{k-1})}{\sqrt{t_k-t_{k-1}}}$, then the random variables $Z_1, Z_2,\ldots,Z_n$ are independent and identically distributed, each with standard normal distribution. This fact is used for the simulation of Wiener processes and SDEs.

Finally, a $d$-dimensional Wiener process  is a collection of $d$ independent one-dimensional
Wiener processes.

\subsection{Martingales and change of measure}\label{Martingale}
Let $\big(\Omega, \Fscr, (\Fscr_t),\mathbb{P}\big)$ be a stochastic basis as defined at the beginning of Section \ref{sec:lSSE}.

An \emph{adapted process} $\{X(t)\}_{t\geq 0}$ is a stochastic process in the probability space $(\Omega, \Fscr, \mathbb{P})$, such that $X(t)$ is $\Fscr_t$-measurable, $\forall t\geq 0$.

A stochastic process $\{X(t)\}_{t\geq 0}$ is said to be a \emph{martingale} if (a) it is adapted, (b) $\Ebb[\abs{X(t)}]<+\infty$, $\forall t \geq 0$, (c) $\Ebb[X(t)|\Fscr_s]=X(s)$, $\forall t \geq s\geq0$.

An adapted Wiener process is a martingale.

Let $Z=\{Z(t), t\geq 0\}$ be a non-negative martingale with $\mathbb{E}[Z(t)]=1$. For every fixed $t\geq 0$, the random variable $Z(t)$ can be used as a density to define a new probability measure $\mathbb{Q}_t$ on $(\Omega,\Fscr_t)$:
\begin{equation}
\forall F\in \Fscr_t\qquad \mathbb{Q}_t(F):=\int_F Z(t,\omega)\mathbb{P}(\rd\omega)\equiv \mathbb{E}[Z(t)1_F].\nonumber
\end{equation}
Being $Z$ a martingale, all the probabilities $\mathbb{Q}_t$, $t\geq 0$, are consistent, in the sense that
 \begin{equation}
 \mathbb{Q}_t(F)=\mathbb{Q}_s(F),\qquad \forall t,s:t\geq s\geq 0,\quad \forall F\in \Fscr_s\nonumber.
 \end{equation}
 Indeed, $1_F$ is $\Fscr_s$-measurable and, by the properties of conditional expectations, one has
\[ 
\mathbb{Q}_t(F)=\mathbb{E}[Z(t)1_F]=\mathbb{E}\left[\mathbb{E}[Z(t)1_F|\mathcal{F}_s]\right]
=\mathbb{E}\left[\mathbb{E}[Z(t)|\Fscr_s]1_F\right]=\mathbb{E}[Z(s)1_F]=\mathbb{Q}_s(F).
\] 

\subsection{Stochastic integrals}\label{Int}
Let $\big(\Omega, \Fscr, (\Fscr_t),\mathbb{P}\big)$ be a stochastic basis, $W$ an adapted Wiener process and $F$ a continuous, adapted, stochastic process with $\Ebb[\abs{F(t)}^2]<+\infty$, $\forall t\in [0,T]$. Then, it is possible to define the \textit{It\^{o} integral}
\begin{equation}\label{integral}
Y(T)=\int_{0}^{T}F(t)\rd W(t)
\end{equation}
as the mean square limit for $\Delta t \downarrow 0$ of
\begin{equation}\label{sum1}
Y_{\Delta t}(T)=\sum_{k=1}^{n-1} F(t_k)\big(W(t_{k+1})-W(t_k)\big),
\end{equation}
where $0=t_0<t_1<\cdots<t_n=T$ is a partition of $[0,T]$ and $\Delta t=\max_k\{t_{k+1}-t_k\}$. This means
\begin{equation}
\lim_{\Delta t\downarrow 0}\Ebb\left[\abs{Y_{\Delta t}(T)-Y(T)}^2\right]=0.
\end{equation}
By approximation techniques, the definition of the stochastic integral can be generalized to an integrand $F(t)$ such that it is adapted and $\int_0^T \Ebb\left[ \abs{F(t)}^2\right] \rd t < +\infty$.

Let us consider the stochastic integral as a process $Y=\{Y(t),\, t\in[0,T]\}$. The main properties of \emph{the integral process} are that it \emph{is a martingale with vanishing mean}, $\Ebb[Y(t)]=0$, and that \emph{the It\^o isometry holds}:
\begin{equation}
 \Ebb\left[\abs{Y(t)}^2\right]=\int_0^t \Ebb\left[ \abs{F(s)}^2\right] \rd s.
 \end{equation}
These properties are easily proved on the discrete approximation \eqref{sum1} and then it is possible to show that they survive to the limiting procedure.

The definition of stochastic integral can be extended to a larger class of integrands (now limits in probability have to be used), but it is no more guaranteed that the main properties hold; we can only say that the integral process is a \emph{local martingale}.

Other definitions of stochastic integral are possible, in particular the Stratonovich integral, whose definition starts from the discrete approximation 
\begin{displaymath}\sum_{k=1}^{n-1} F\big((t_k+t_{k+1})/2\big)\big(W(t_{k+1})-W(t_k)\big).\end{displaymath} While the rules of the stochastic calculus based on the Stratonovich definition are simpler than the ones based on It\^o integral, the important properties above are lost.

\subsection{It\^{o} calculus}\label{Calc}
Let now $W$ be a $d$-dimensional Wiener process defined in the stochastic basis $(\Omega, \Fscr, (\Fscr_t), \Pbb)$. An \emph{It\^o process} $X$ is a continuous, adapted process such that $X(0)$ is $\Fscr_0$-measurable and
\begin{equation*}
X(t)= X(0) +\int_0^t F(s) \,\rd s +\sum_{j=1}^d \int_0^t G_j(s)\,\rd W_j(s),
\end{equation*}
for some adapted process,  $F$ Lebesgue integrable and $G_j$ stochastically integrable. It is usual to say that $X$ admits the \emph{stochastic differential}
\begin{equation}\label{dX(t)}
\rd X(t)= F(t) \,\rd t + \sum_{j=1}^d  G_j(t)\,\rd W_t(t).
\end{equation}

Take now another It\^o process with stochastic differential
\begin{equation}\label{dY(t)}
\rd Y(t)= M(t) \,\rd t + \sum_{j=1}^d  N_j(t)\,\rd W_t(t).
\end{equation}
The \emph{It\^o lemma} says that the product $X(t)Y(t)$ of two It\^o processes is an It\^o process with initial value $X(0)Y(0)$ and stochastic differential
\[
\rd \big(X(t)Y(t)\big)=X(t)\, \rd Y(t)+ Y(t)\, \rd X(t) + \bigl(\rd X(t)\bigr) \bigl(\rd Y(t)\bigr),
\]
where $\rd X(t)$, $\rd Y(t)$ have the expressions \eqref{dX(t)}, \eqref{dY(t)}, and the \emph{It\^o correction} $\bigl(\rd X(t)\bigr) \bigl(\rd Y(t)\bigr)$ must be computed from the product of the two differentials by using the \emph{It\^o table}
\[
(\rd t)^2=0, \qquad \rd t\, \rd W_j(t)=0, \qquad \rd W_j(t)\, \rd W_i(t) = \delta_{ij} \, \rd t.
\]

This result can be generalized to polynomials in $W$ and then to smooth functions of $W$; this is the It\^o formula \cite{KarS91,Mao97}, \cite[Sections A.3.3, A.3.4]{3}.

\end{document}